# Characterizing a CCD detector for astronomical purposes: OAUNI Project

# Caracterizando un detector CCD para uso astronómico: proyecto OAUNI

**Antonio Pereyra[1*], María I. Zevallos[2,3], José Ricra[2], Julio C.Tello[2]**

[1] Instituto Geofísico del Perú, Área Astronomía, Calle Badajoz169, Mayorazgo, Lima, Perú

[2] Grupo Astronomía, Facultad de Ciencias, Universidad Nacional de Ingeniería, Av Túpac Amaru. 210, Rimac Lima, Perú

[3] Facultad de Ingeniería Química y Textil, Universidad Nacional de Ingeniería, Av Túpac Amaru. 210, Rimac Lima, Perú

**ABSTRACT**

This work verifies the instrumental characteristics of the CCD detector which is part of the UNI astronomical observatory. We measured the linearity of the CCD detector of the SBIG STXL6303E camera, along with the associated gain and readout noise. The linear response to the incident light of the detector is extremely linear ($R^2$ =99.99%), its effective gain is 1.65 ± 0.01 e-/ADU and its readout noise is 12.2 e-. These values are in agreement with the manufacturer. We confirm that this detector is extremely precise to make measurements for astronomical purposes.

**RESUMEN**

El presente trabajo verificó las capacidades del detector CCD que forma parte del instrumental asociado al proyecto del Observatorio Astronómico de la UNI. Fueron medidos la respuesta lineal del detector CCD de la cámara SBIG STXL6303E, así como su ganancia y el ruido de lectura asociados. La respuesta lineal a la luz incidente del detector es extremadamente lineal ($R2$=99.99%), su ganancia efectiva es 1.65 ± 0.1 e-/ADU y su ruido de lectura 12.2 e-. Estos valores están de acuerdo con los proporcionados por el fabricante. Se confirma que el detector analizado es extremadamente preciso para realizar medidas con fines astronómicos.

*Palabras clave: Dispositivos de carga acoplada, Linealidad, ganancia, Ruido de lectura*

## 1. INTRODUCTION

The National University of Engineering (UNI) has an astronomical observatory project (OAUNI, [1]) at the peruvian central Andes (Huancayo, 3300 m.u.s.l.). This ongoing effort aims to provide of a facility to develop science programs, teaching and outreach in astronomy. The observatory has several instruments being the most important the scope and the detector. The proper selection of both is necessary to acquire astronomical images of quality.

CCD detectors are the standard device to register optical digital images in practically all types of applications ranging from domestic and scientific ones. They are bidimention arrays of detection elements (or pixels) that convert incident photons on them into electrons. In particular, their use in professional astronomy let a huge enhancement on the precision of photometric and spectroscopic measurements of astronomical objects over the last decades.

The great advantage of CCDs when comparing with other detectors (as photographic emulsions, for example) is its linearity to the response of incident light. In addition, the better sensibility for the optical spectral range makes the CCDs the natural choice for astronomical applications ([2]). There are several types of CCDs with different approaches (front-illuminated, back-illuminated, full frame, frame transfer, etc.; [3],[4]) to enhance the sensibility and linearity depending on the particular application.

In general, each CCD is characterized by its linearity, quantum efficiency (sensibility), gain and readout noise. The gain refers to the conversion between the number of electrons (e-) recorded by the CCD and the number of analog to digital units

* Correspondencia:
Email: apereyra@igp.gob.pe

(ADU) contained in the CCD image. Gain is given in (e-/ADU).

The three main sources of noise in CCD measurements are the readout noise, the thermal noise and the photon noise. The readout noise (R) is present in all images with the same amount regardless of integration time. This represents the on-chip noise source that affect the measurement and it is given in e- RMS.

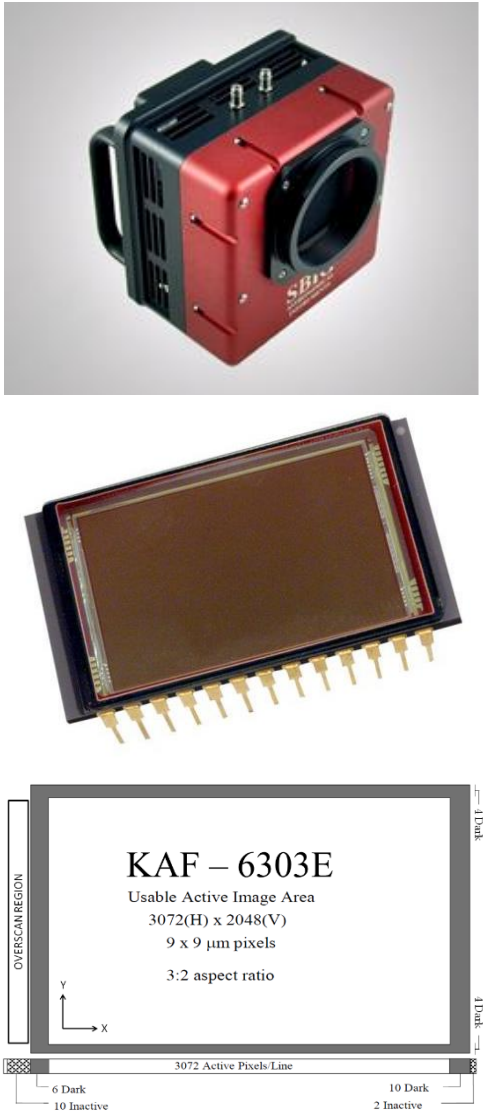

Fig.1 (top) SBIG STXL-6303E Camera [5]; (middle) CCD Chip KAF-6303; (bottom) chip diagram (adapted from [6]).

The thermal noise (ND) is dependent of the temperature of the chip and responds to electrons (D) created over time that are independent of the light falling on the detector. It can be reduced at lower temperatures. Finally, the photon noise (N) depends on the amount of light hitting the chip (S). The last two noises responds to the Poisson statistics ($N_D^2 = D$, $N^2 = S$). All the noises given in electrons are added in quadrature to the total noise, as following.

$$N^2_{tot} = N^2 + N_D^2 + R^2, \quad \text{or,}$$

$$= S + D + R^2 \quad (1)$$

The CCD camera of OAUNI's project is of the extreme sensibility and its characterization is needed in order to be used properly on the observational programs which include photometry and spectroscopy. The camera is a SBIG STXL6303E (Fig. 1, top) from Diffraction Limited / SBIG provider and it includes the front-illuminated CCD chip Kodak KAF-6303E (Fig. 1, middle).

This work presents systematic laboratory tests in order to characterize the linearity, the gain and readout noise of the CCD KAF 6303E. These values are then compared with the specifications given by the manufaturer. Finally, the conclusions about the feasibility of the OAUNI's detector to make precise astronomical measurements are summarized.

## 2.   MEASUREMENTS

Measurements using the camera STXL-6303E were performed in controlled conditions. They consisted in sequences of flat field and dark current images with a CCD temperature of -5°C at several integration times. The image acquisition was done using the software CCDOps ver. 5.55. The flat field images were used to compute the linearity and gain of the CCD.  On the other hand, the dark current images were used to calculate the readout noise. A white LED source was used to obtain a homogenous illumination of the roof lab for flats fields. The camera was located pointed to the roof with a proper lateral protection to avoid reflected light. In order to obtain larger integration times, an astronomical broadband blue filter was used in front of CCD. The effective central wavelength for this filter is 4353Å with a broadband of 781Å. The attenuation was enough to perform integration times (IT) between 1 and 14 seconds (in steps of 0.5s) covering all the dynamical range of the CCD. A short shoot of 0.4s also was gathered. For each IT, 5 different shoots were done. The 4.0s sequence was incidentally lost and it was not considered on the below analysis.

Dark current images were acquired with the lights off and the entrance of the camera closed. Again, 5 shoots by IT were performed for 0.4, 1, 2, 3, 5, 8, 11 and 14s.

All the images were acquired including the overscan mode in CCDOps software. This let include and measure the electronic offset associated with the readout process. This is done on each image in a proper section of inactive pixels. This section is defined by the manufacturer and for the CCD KAF-6303E this is indicated in Fig 1. (bottom).

## 3.   REDUCTION

All the images were reduced using the IRAF 2.16 software in Linux/Ubuntu 12.04 operational system. The first step was to perform the overscan correction and to select the optimal region for the data (or trimming section). These sections are indicated in Table 1. Flat field and dark images were then corrected by overscan subtracting the mean value computed on the overscan region for each individual image. Fig. 2 shows a histogram of a typical overscan region. The mean value is aprox. 1000 ADU. After overscan correction, cropping was applied using the trimmed section to recover the operational working area of the chip in each image.

Table 1. KAF-6303E regions.

| section | [X$_{begin}$:X$_{end}$,Y$_{begin}$:Y$_{end}$] |
|---------|-----------------------------------------------|
| full chip | [ 1:3100, 1:2056] |
| overscan | [ 1:10  ,50:2010] |
| trimmimg | [50:3050,50:2010] |

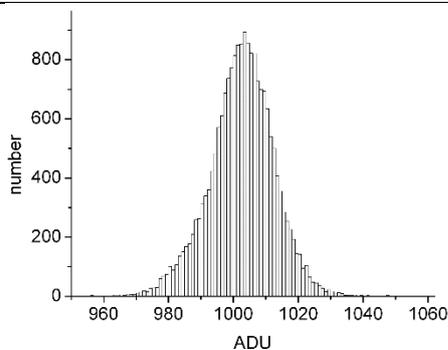

Fig 2. Histogram for the overscan region. The mean value is 1002.00 ± 10.08 ADU. The error is the sample standard deviation.

In principle, a zero bias level may still prevail after the overscan correction. This zero level calibration is obtained by taking images with zero integration times. In order words, this level considers the electronic offset on active pixels. Several zero bias images must be taken and then properly combined to get an averaged zero bias image. In our case, the CCDOps software only lets 0.4s as the minimum integration time. We used the 0.4s dark current images, after overscan correction, as representative of the zero bias level. The sequence of 5 images was averaged and Fig. 3 shows the result. A residual mean bias value of ~11 ADU is detected.

After that, all the flat field and darks images were corrected subtracting this mean zero bias in each image. The 5-images sequences for the dark current images were then combined and averaged for each integration time.

For test the linearity, two approaches were used with the flat fields. The first was including only the mean zero bias correction. The second one was consider only dark current correction. This let us to know if the temperature dependence of the dark current is important or not on the linearity of the detector for the temperature used in these tests.

With this, each 5-images sequence of flat fields for each type of correction was properly combined and averaged.

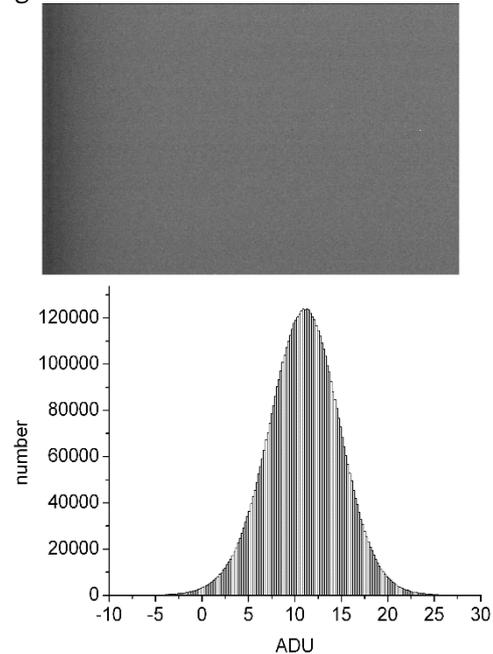

Fig 3. (top) Mean zero bias image. (bottom) Histogram of mean zero bias image. The mean value along the chip is 11.04 ± 3.95 ADU. The error is the sample.

## 4. RESULTS

The test of linearity [7] was performed in a homogenous subsection of the flat field images (as seen in Fig 4). The mode for the pixels in this subsection of the chip was computed for each averaged flat image and in each integration time available. This step was done for both flat field corrections commented above. The results are shown in Table 2. Only 8 points are available with the dark current correction because this is the same number of darks sequence available.

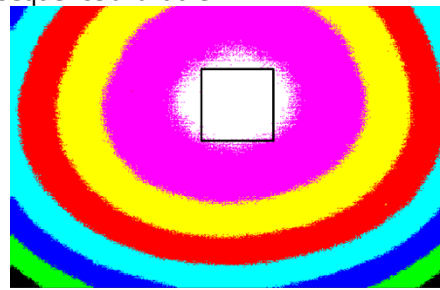

Fig. 4. Averaged flat field image (for IT=14s) corrected by overscan and dark current in false color to highlight the typical pattern. The black box indicates the 500x500 pixel subsection where the mode of the signal was computed for the linearity tests.

The linearity of the chip KAF-6303E including the bias correction is plotted in Fig. 5 The dynamical range sampled goes from 2k to 61k ADU. The reference full well capacity for this chip is 100k e-. Considering the nominal gain of 1.47 e-/ADU, our fit ranges between 3 and 90% of the full well capacity. The coefficient of determination (CoD) for the linear fit is R2 = 99.998%, indicating an extreme linearity of the chip.

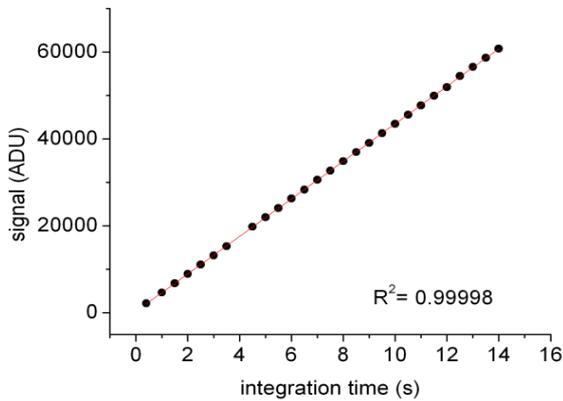

Fig. 5 Linearity tests for CCD KAF-6303E detector using flat field images for several integration times. Data include overscan and bias correction. Mean values for a regular section of 500x500 pixels are shown (black dots). The coefficient of determination for the linear fit (red) is also shown (R2= 99.998%).

The linear fit for the flat fields corrected by dark current is plotted in Fig. 6. Again, the CoD (R² = 99.999%) indicates a sharp linearity for the chip. The effect of the dark current correction does not seem important for the temperature used in this tests (-5C). In principle, scientific data can be reduced only with overscan and bias correction when dark current frames are not available. Of course, flat fielding also is necessary for a complete calibration process.

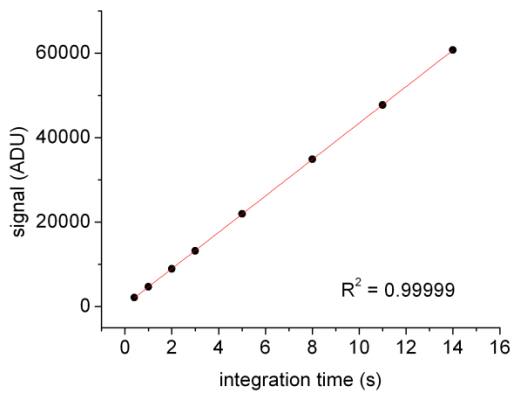

Fig. 6 As Fig. 1 including only overscan and darks current correction (R2 = 99.999%).

As shown in eq. 1, the signal (S) on each pixel can be represented by the Poisson statistic after correction by the readout and dark current noises, following,

$$N^2 = S$$

This can be applied when the signal and the noise are counted in electrons. As the gain ($g$) relates electrons and ADUs, the above expression can be expressed by,

$$(gN')^2 = gS' \qquad , \text{or,}$$
$$N'^2 = (1/g)\, S'$$

where $N'$ and $S'$ are in ADUs. Therefore, measuring the signal and its noise (both in ADUs) for different signal levels (or integration times, in our case), the inverse of the slope will be the gain.

In order to get an estimate of the gain [8] we compute the average and the difference images of two flats with the same integration time,

$$avg_{IT} = (\, flat1_{IT} + flat2_{IT}\,)\, /\, 2$$
$$dif_{IT} = (\, flat1_{IT} - flat2_{IT}\,)$$

We used the flat fields corrected by dark current in this exercise. The statistics were made also in the same subsection used for the linearity tests. The average image represents the signal ($S'$),

$$S' = avg_{IT}$$

and the variance (or the square of the standard deviation) divided by 2 of the difIT image is associated to the squared signal noise ($N'^2$),

$$\begin{aligned} N'^2 &= \text{variance } (dif_{IT}) \,/\, 2 \\ &= \text{stddev}^2 \,(dif_{IT})\,/\,2. \end{aligned}$$

Table 2. Flat fields statistics. The signal is measured by the mode.

| IT | w/ bias correct. | w/ dark correct. |
|---|---|---|
| (s) | (ADU) | (ADU) |
| 0.4 | 2146 | 2146 |
| 1.0 | 4663 | 4664 |
| 1.5 | 6777 | -- |
| 2.0 | 8928 | 8917 |
| 2.5 | 11070 | -- |
| 3.0 | 13183 | 13194 |
| 3.5 | 15293 | -- |
| 4.5 | 19795 | -- |
| 5.0 | 21966 | 21962 |
| 5.5 | 24073 | -- |
| 6.0 | 26292 | -- |
| 6.5 | 28327 | -- |
| 7.0 | 30584 | -- |
| 7.5 | 32680 | -- |
| 8.0 | 34892 | 34887 |
| 8.5 | 36985 | -- |
| 9.0 | 39067 | -- |
| 9.5 | 41289 | -- |
| 10.0 | 43483 | -- |
| 10.5 | 45601 | -- |
| 11.0 | 47716 | 47720 |
| 11.5 | 49905 | -- |
| 12.0 | 51862 | -- |

| 12.5 | 54452 | -- |
|------|-------|-----|
| 13.0 | 56600 | -- |
| 13.5 | 58635 | -- |
| 14.0 | 60730 | 60736 |

Table 3. Gain factor statistics.

| IT | $S$ | $N'^2$ |
|-----|-------|---------|
| (s) | (ADU) | (ADU$^2$) |
| 0.4 | 2145 | 1601 |
| 1.0 | 4657 | 3183 |
| 2.0 | 8904 | 5832 |
| 3.0 | 13181 | 8476 |
| 5.0 | 21957 | 13894 |
| 8.0 | 34891 | 21882 |
| 11.0 | 47723 | 29573 |
| 14.0 | 60856 | 36856 |

The final noise is computed on the difIT image because the subtraction eliminates all the noise sources except the Poisson noise. The statistic is indicated in Table 3. The slope of the linear fit in Fig. 7 yields the gain for the chip (g = 1.654 ± 0.012 e-/ADU). The nominal gain as appears on the specifications of the manufacturer is 1.47 e-/ADU and the difference with respect to our value is 12%. Our tests are in a reasonable agreement with the nominal gain.

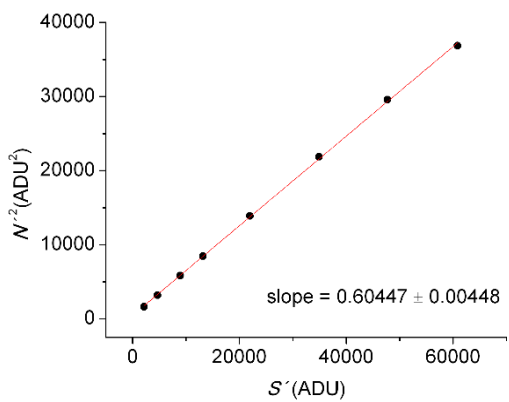

Fig. 7. Gain for the CCD KAF-6303E. The gain computed from the inverse of the slope is 1.654 ± 0.012 e-/ADU.

In order to compute the readout noise of our chip, we used the dark current images obtained at several integrations times [9]. Five readout noise images were constructed for each IT subtracting of the averaged dark current image each individual dark frame, following,

$$RON_{IT}{}^i = dark_{IT}{}^{avg} - dark_{IT}{}^i$$

A total of (8x5=) 40 RON images were computed. As each dark current image only includes the thermal signal without incident light on the chip, the subtraction of the averaged image results in only one term for the total noise in eq. 1,

$$N^2{}_{tot} = R^2.$$

With this, the mean value of each RON image is of course close to zero but its standard deviation represents the readout noise. The statistics was made considering the full frame.

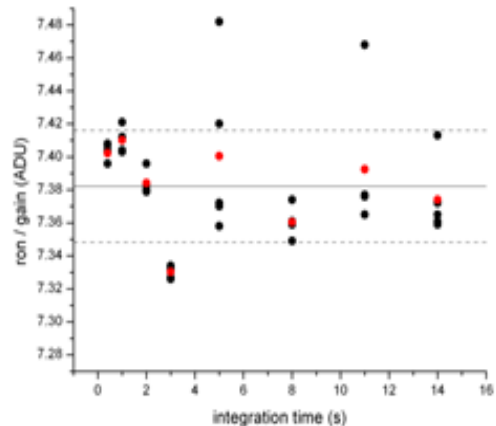

Fig. 8. Computing of readout noise for CCD KAF-6303E. The standard deviation of the differences between individual darks (without overscan correction) and their mean average images are shown for several integration times(black dots). The mean values for each integration time are also shown (red dots). The mean standard deviation (7.382 ± 0.034 ADU) for all the measurements is indicated (solid line) along with its respective one-sigma dispersion (dotted lines). With the gain value (1.654 e-/ADU) computed before, the mean readout noise is 12.209 e-.

Fig. 8 shows the readout noise computed for all the dark current images following the above procedure using images without overscan correction. Considering the 40 RON images, the mean readout noise value is 7.382 ± 0.034 ADU. The dispersion is small and the result is very precise. The same exercise using dark current images with overscan correction yields a similar result.

In order to compare with the readout noise given by the manufacturer (11e-, [9]), we convert this value to ADU using the proper manufacturer's gain (11e- / (1.47 e-/ADU) = 7.48 ADU). The accuracy of our result is evident. Finally, if we transform our RON result using our own gain value computed before, we obtained RON = 12.2 e-. This compares very well with the manufacturer value being the difference only of 11%.

## 5. CONCLUSIONS

We present calibration tests of the SBIG STXL-6303E camera. These include a verification of the main specifications of the CCD chip as linearity, gain and readout noise. Our tests indicate a rigorously linear detector (R2 = 99.99%) between 3 and 90% of its full well capacity. The gain and readout noise computed in this work are slightly higher than the

                    

manufacturer values in ~11-12%. The gain computed was 1.654 ± 0.012 e-/ADU and the readout noise was 12.2 e-. In general, these values are in agreement with the nominal values for this detector. The performed calibrations tests show the feasibility of the chip KAF-6303E to make precise measurements for astronomical purposes.

## ACKNOWLEDGEMENTS

The authors are grateful for the economic support from The World Academy of Sciences (TWAS) and the Instituto General de Investigación (IGI) at UNI.